\documentclass{aa}
\usepackage{graphicx}

\def\gtsima{$\; \buildrel > \over \sim \;$}
\def\ltsima{$\; \buildrel < \over \sim \;$}
\def\prosima{$\; \buildrel \propto \over \sim \;$}
\def\gsim{\lower.5ex\hbox{\gtsima}}
\def\lsim{\lower.5ex\hbox{\ltsima}}
\def\simgt{\lower.5ex\hbox{\gtsima}}
\def\simlt{\lower.5ex\hbox{\ltsima}}
\def\simpr{\lower.5ex\hbox{\prosima}}

\def\h1{$h^{-1}$}

\def\beq{\begin{equation}}
\def\eeq{\end{equation}}
\begin{document}
   \title{The K20 survey. II. The Different Spatial Clustering of $z\sim1$\\ Old and Dusty Star-Forming EROs\thanks{Based on observations made at the
   European Southern Observatory, Paranal, Chile (ESO LP 164.O-0560)}}

   \author{E. Daddi\inst{1,10}
          \and A. Cimatti\inst{2}
	  \and T. Broadhurst\inst{3}
  	  \and A. Renzini\inst{1}
	  \and G. Zamorani\inst{4}
	  \and M. Mignoli \inst{4}
	  \and P. Saracco \inst{5}
	  \and A. Fontana \inst{6}
	  \and L. Pozzetti \inst{4}
	  \and F. Poli \inst{7}
	  \and S. Cristiani \inst{8,9}
	  \and S. D'Odorico \inst{1}
	  \and E. Giallongo \inst{6}
	  \and R. Gilmozzi \inst{1}
	  \and N. Menci \inst{4}
          }

   \offprints{{edaddi@eso.org}}
\authorrunning{E. Daddi et al.}
\titlerunning{3D Clustering of EROs in the K20 survey} 

   \institute{
   European Southern Observatory, Karl-Schwarzschild-Str. 2, D-85748,
   Garching, Germany
   \and Osservatorio Astrofisico di Arcetri, Largo E. Fermi 5, I-50125,
   Firenze, Italy
   \and Racah Institute for Physics, The Hebrew University, Jerusalem,
   91904, Israel
   \and Osservatorio Astronomico di Bologna, via Ranzani 1, I-40127, Bologna, Italy
   \and Osservatorio Astronomico di Brera, via E. Bianchi 46, Merate,
   Italy
   \and Osservatorio Astronomico di Roma, via Dell'Osservatorio 2,
   Monteporzio,
   Italy
   \and Dipartimento di Astronomia, Universit\`a ``La Sapienza'', Roma,
   Italy
   \and ST, European Coordinating Facility, Karl-Schwarzschild-Str. 2,
   D-85748,
   Garching, Germany
   \and Osservatorio Astronomico di Trieste, via Tiepolo 11, I-34131,
   Trieste, Italy
   \and Dipartimento di Astronomia, Universit\`a di Firenze, Largo E.
   Fermi 5, I-50125, Firenze, Italy
   }

   \date{Received; Accepted}

   \abstract{We compare the 3D clustering of 
    old passively-evolving  and dusty star-forming $z\sim1$ 
    EROs from the K20 survey. 
    With detailed simulations of clustering, the
    comoving correlation length of dusty star-forming EROs is constrained
    to be less than $r_0 \sim 2.5$ \h1 Mpc. 
    In contrast, the old EROs are much more positively
    correlated, with $5.5 \simlt r_0/($\h1 Mpc$) \simlt 16$, 
    consistent with previous claims for $z\sim1$ field early-type galaxies
    based on analyses of ERO angular clustering. 
    The low level of clustering of dusty 
    star-forming EROs does not support
    these to be major mergers building up an elliptical
    galaxy, or typical counterparts of SCUBA sources, but it is
    instead consistent with the weak clustering of high redshift blue
    galaxies and of luminous local IRAS galaxies. 
    Current hierarchical merging 
    models can explain the large $r_0$ for $z\sim1$
    field early-type galaxies, but fail in matching their high number density
    and overall old ages.
\keywords{Galaxies: evolution; Galaxies: elliptical and lenticular,
   cD; Galaxies: starburst; Galaxies: formation; large-scale structure
   of Universe}}

   \maketitle

\section{Introduction}

Extremely red objects ($R-K>5$, EROs hereafter) are providing
increasingly stringent constraints on our understanding of the formation
of galaxies in general, via their spectral evolution and clustering
properties. The very red colors of EROs are well known to be both
consistent with old passively evolving distant ($z>0.8$) elliptical
galaxies (e.g. Cohen et al. 1999; Spinrad et al. 1997) or dust-reddened
starburst galaxies (e.g. Cimatti et al. 1998; Smail et al. 1999). 
Purely passive evolution of the present day population of
elliptical galaxies is consistent with the measured surface density of
faint EROs with $K\sim17$--22, while  
current renditions of the semianalytical hierarchical merging models
fail to reproduce the surface density of EROs by a large factor
(Daddi et al. 2000a; Smith et al. 2001; Firth et al. 2001).

Recently, we have completed a relatively large deep survey of very red
galaxies covering 700 arcmin$^2$ (Daddi et al. 2000b, D00
hereafter), concluding that EROs are strongly clustered in projection,
by an order of magnitude more than all galaxies at the same
limits of $K\leq 18$--19.2. With careful attention to the measurement
uncertainty inherent in narrow field data, Daddi et al. (2001, D01
hereafter) showed that the angular clustering of EROs implies a
spatial correlation length of $r_0=12\pm3$ \h1 comoving Mpc, 
consistent with the
assumption that the ERO population is dominated by elliptical
galaxies. This large clustering amplitude is not at odds with
recent hierarchical merging models, which require that the most
massive galaxies are clustered more strongly than the general galaxy
population at high $z$ (e.g. Mo \& White 1996). 
Our results on the angular and spatial clustering of EROs
have been substantially confirmed by the Las Campanas Redshift Survey
data (McCarthy et al. 2001; Firth et al. 2001; Moustakas \&
Somerville 2001).

{In our recent large K20 redshift survey of a flux limited sample of
$\sim500$ galaxies with $K\leq20$} 
(Cimatti et al. 2002, C02 hereafter), we
obtained redshifts for a sub-sample of 35 EROs. For
red objects with $R-K>5$ and $K\leq19.2$, about 1/3 were
identified as old systems (consistent with being passively evolving
elliptical galaxies), 1/3 were found to be dusty starburst galaxies and
1/3 remain unidentified.
While the derived fraction of early-type galaxies, $50\pm20$\%, 
is consistent with previous estimates based on morphology 
(Moriondo et al. 2000; Stiavelli \&
Treu 2000), C02 showed that the dusty star-forming (SF) objects do contribute
significantly to the ERO population at faint magnitudes, thus
complicating the interpretation of both ERO surface density and 
clustering, as measured in earlier analyses.  In particular,
given the strong interest in the clustering amplitude of
early-type galaxies, it is important 
to estimate separately the clustering properties of the old and of the 
dusty-SF EROs, hence their relative contribution to the clustering of
the whole ERO population.
This is attempted in this letter, where we adopt a 
cosmology with $\Omega_\Lambda = 0.7$, $\Omega_m = 0.3$ and
$H_0 = 100h$ km/s/Mpc.

\section{Clustering analysis}

\subsection{The sample and the diagnostic method}

Table \ref{tab:C02} shows the redshifts of the EROs identified in the
K20 survey (C02) and classified as old passively evolving 
or dusty-SF galaxies,
sorted with increasing redshift and divided between the two survey
fields (32.2 arcmin$^2$ from CDFS and 19.8 arcmin$^2$ from 0055-27). 
{The classification of EROs as old galaxies is based on the detection
of the 4000\AA\ break and CaII H\&K absorption with undetected (or very weak) 
[OII]$\lambda$3727 emission, while objects with strong
[OII]$\lambda$3727 emission and an absence of a distinctive 4000\AA\
break were assigned to the dusty-SF  class (see C02 for details).}

Despite being by far the largest sample of EROs with identified redshifts,
standard methods for evaluating the full two
point correlation function cannot be still applied because of the small number
of objects.  Nevertheless the
clustering properties of the old and dusty-SF samples can be investigated
by studying the frequency of close pairs.  This kind of approach has
been applied in regimes with limited amount of information, e.g. 
to early studies of QSO clustering (Shaver 1984,
cfr. also Hartwick \& Schade 1990), or to analyses of
the arrival directions of ultra high energy cosmic rays (Tinyakov \&
Tkachev 2001), 
and relates to the integral under the correlation function on small
scales, where most of the amplitude lies.

\begin{table}[ht]
\begin{flushleft}
\caption{ERO redshifts in the K20 survey. 
All but four EROs have $K\leq19.2$.
The redshift measurement errors are preliminarily estimated to
be of the order of $\sigma\sim100$--200 km/s.
}
\protect\label{tab:C02}
\begin{tabular}{cc|cc}
\noalign{\smallskip}
\hline
\noalign{\smallskip}
\multicolumn{2}{c}{CDFS} & \multicolumn{2}{c}{0055-27} \\
\noalign{\smallskip}
\multicolumn{1}{c}{\em Old} & \multicolumn{1}{c}{\em Dusty-SF} &
\multicolumn{1}{c}{\em Old} & \multicolumn{1}{c}{\em Dusty-SF} \\
\noalign{\smallskip}
\hline
\noalign{\smallskip}
0.726 & 0.796 & 0.790 & 0.820 \\
1.019$^1$ & 0.863 & 0.864 & 0.996 \\
1.039 & 0.891 & 0.896 & 1.210 \\
1.096 & 0.974 & 0.896 & 1.240 \\
1.215 & 0.996$^1$ & 0.935 & 1.300$^1$ \\
1.222 & 1.030 & 1.050 & 1.419 \\
1.222 & 1.094 & 1.104 & \\
 & 1.109$^1$ & 1.166 & \\
 & 1.149 & & \\
 & 1.221 & & \\
 & 1.294 & & \\
 & 1.327 & & \\
\noalign{\smallskip}
\hline
\end{tabular}
\end{flushleft}
\footnotesize{$^1$ Objects with $19.2<K\leq20$}
\end{table}

From Table \ref{tab:C02}, it can be noted that the sample of
old EROs contains two pairs that, within the observational
redshift accuracy, have the same redshift ($z=0.896$
in the 0055-27 field and $z=1.222$ in the CDFS), with an additional
object close to the second pair at $z=1.215$. On the other hand, the
sample of dusty-SF EROs contains no really close pair, the closest pair
having a relatively large redshift
separation $\Delta z = 0.015$ ($z=1.094$ and $z=1.109$ in the CDFS,
{corresponding to $\Delta v\sim 4500$ km/s).}
The two old ERO pairs with the same redshift have also quite small
angular separations ($\simlt 1\arcmin$), implying
spatial separations
of $0.51$ and $0.82$ \h1 Mpc,
while the two closest dusty-SF pairs are separated by
24 and 40 \h1 Mpc, respectively.  The number of independent pairs in the samples is 81 for
the dusty-SF EROs and 49 for the old EROs, thus immediately
suggesting a higher intrinsic clustering amplitude 
for the old EROs.

To assess the significance of observed pair counts we first generate 
random samples. The selection functions are constructed
from the observed redshift distributions of the two ERO
populations. 
Simulated samples were built by assigning at random a redshift 
(rounded to $\Delta z=0.001$ to match the data
redshift measurements) extracted
from the appropriate selection function, with
sky positions within boundaries matching the area of each
of our fields, and number of objects as in the relative observations (Table 1). 
The resulting probability of finding by
chance $\geq2$ pairs of old EROs within a separation $\leq0.82$ \h1
Mpc is about $5\times 10^{-5}$, a clear
evidence of clustering among the sample of old EROs. On the
other hand,
the probabilities of finding the closest dusty-SF ERO pair 
at $\leq24$ \h1 Mpc and the
two closest pairs at $\leq40$ \h1 Mpc are both $\sim97\%$, 
consistent with purely random chance.  

\begin{figure}[ht]
\centering
\includegraphics[width=9cm]{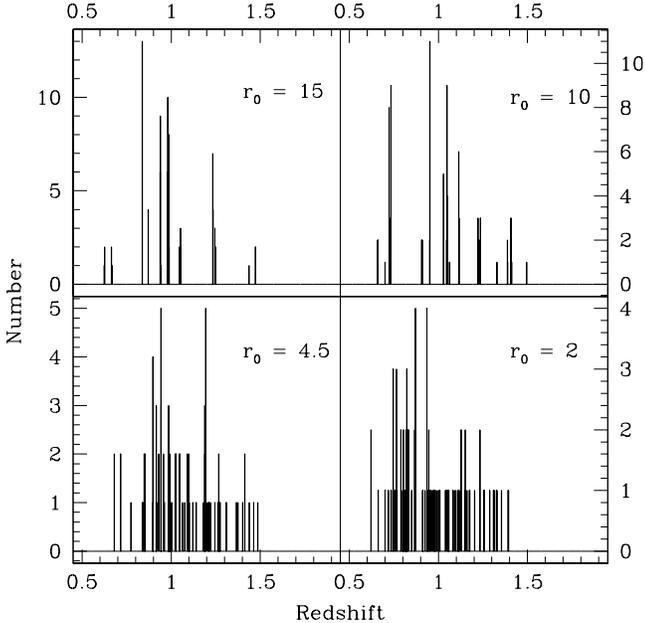}
\caption{
For each panel we show the redshift distribution 
of 100 objects extracted from 
our simulations,  
that incorporates
a given correlation length $r_0$, as indicated on the figure, illustrating
the influence of the correlation amplitude on the overall smoothness of 
pencil beam redshift surveys (see text).
}
\label{fig:exa}
\end{figure}

\subsection{Comparison to clustered samples}

 We now proceed a step further and generate simulated samples
incorporating a known 2-point clustering amplitude, in order to 
derive information on the clustering of the two classes,
and to obtain
meaningful estimates of the variance inherent in the pairs statistics
in the sample. We follow the recipe described in D01, based on
the Soneira \& Peebles (1977; 1978) prescription, allowing us to generate
many samples with a given value of $r_0$ over very large volumes. We adopt
the canonical parameterisation $\xi(r)\propto r^{-\gamma}$ with a 
slope of $\gamma=1.8$ (justified by the 
observed angular slope $\delta=0.8$, D00) 
for the 2-point correlation function and allow the amplitude to vary.

 For these simulations one has also to account for the
redshift space distortion, which tends to decrease the numbers of small
scale pairs, and for the  measurement error in the redshift. For the 
pairwise peculiar velocity dispersion 
we adopt the local value of $\sigma_{12}=360$ km/s (Landy, Szalay
\& Broadhurst 1998, see also Peacock et al. 2001) 
and their functional parameterisation, which is
assumed not to evolve significantly over the redshift range of our
data (e.g. Kauffmann et al. 1999).  
For the redshift error $\sigma=150$ km/s is adopted (cfr. Table 1),
and we note that
its contribution is small compared to the peculiar velocity term.  To
each simulated object, an error in the redshift measurement and a
peculiar velocity is added in quadrature, chosen randomly from the
appropriate distributions, before rounding its redshift to 
$\Delta z=0.001$ to match the data redshift measurements.

\begin{figure}[ht]
\centering
\includegraphics[width=9cm]{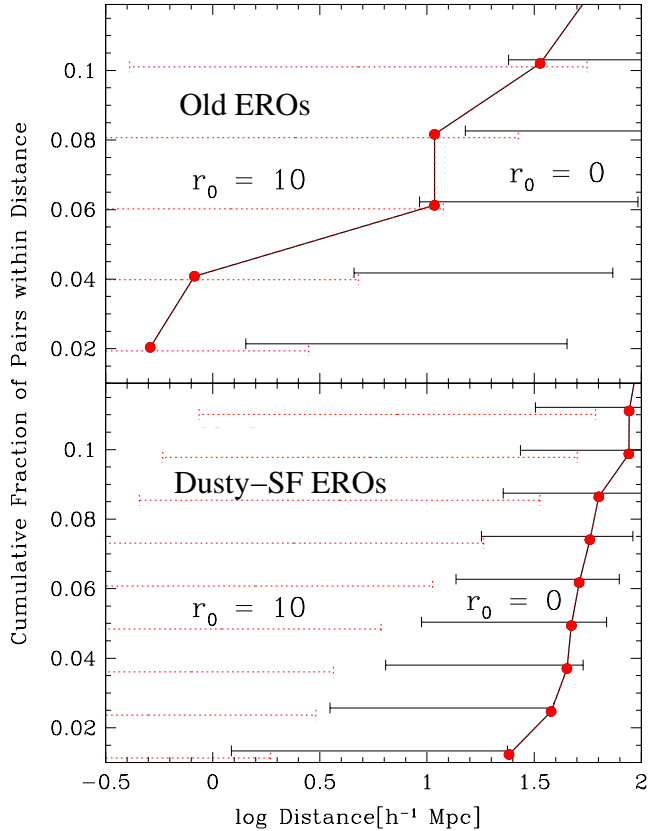}
\caption{ Top panel: the cumulative distribution of pair separations
observed for the old EROs (heavy line with filled circles). The
horizontal error bars show the $2\sigma$ range estimated from our
simulations with random (solid lines) and clustered (dotted lines,
$r_0=10$ \h1 Mpc) realizations. Bottom panel: the same but for the
dusty-SF EROs.  
This comparison shows that while
the error on the  estimate of the correlation length of either sample
is quite broad, it is clear that the dusty-SF EROs as a class are
completely inconsistent with a correlation length of order 10 \h1 Mpc,
estimated from projected samples of EROs (D01, Firth et
al. 2001).  }
\label{fig:huub}
\end{figure}

As expected, the close pairs statistics is strongly
dependent on the correlation length. For example,
Fig.  \ref{fig:exa} shows 
that in the case of strong clustering almost all the objects reside in
spikes with 2 or more objects in each $\Delta z = 0.001$ bin, and
therefore even with our small number of objects we would
expect to find a number of very close pairs (as indeed we do find
for the old EROs).
In fact, with the clustered samples the probability to find $\geq2$
pairs within $\leq0.82$ \h1 Mpc increases strongly with
$r_0$ and at the $1\sigma$ level the observed
close pairs statistics requires the correlation length of the
present sample of old
EROs to lie in the broad range $5.5 \simlt r_0/($\h1 Mpc$)\simlt 16$.  On
the other hand, for the dusty-SF EROs, the observation of the two
closest pairs within 40 \h1 Mpc constrains $r_0<2.5$ \h1 Mpc at the $3\sigma$ 
confidence level.
Fig. \ref{fig:huub} summarises concisely the comparison between the
fraction of observed pairs below a given scale compared with 
the random ($r_0=0$) and clustered ($r_0=10$ \h1 Mpc) expectations for 
a range of scales.

\subsection{Spatial and angular clustering of $K\leq19.2$ EROs}

If we assume $r_0<2.5$ \h1 Mpc for the observed sample of dusty-SF
EROs, this results in an angular clustering amplitude $A(1^o)\simlt 0.002$
at $K\sim19$.  We recall that EROs as a whole (including both
old and dusty-SF objects) have a factor of 10 larger angular
amplitude than this (D00). 
A solid result of this analysis is therefore that the dusty-SF EROs cannot be
the cause of the strong angular clustering of EROs reported by D00, in
agreement with the considerations of D01.  
It is clear from our redshift survey that a significant fraction of
EROs are weakly clustered dusty-SF galaxies which therefore dilutes the
true angular clustering amplitude of the early-type galaxies
population responsible for the majority of the clustering signal.
A detailed estimate of the amplitude of this dilution effect 
would  need a more precise
knowledge of the relative fractions of both classes and a measure of
the cross correlation between the two ERO species.  
{In fact, two dusty-SF EROs are in close redshift pairs with old EROs
(see Table \ref{tab:C02}), 
with $\Delta z
\leq 0.002$ and distances within 3.2 \h1 Mpc, with a probability of
only 2\% to happen by chance. This is evidence of some
positive cross-correlation bewteen the two ERO species, an intriguing 
result considering the different physical properties of the two
populations.
We defer a
discussion of this aspect as a part of the ongoing analysis of the 
clustering of the whole K20 sample (Daddi et al. 2002, in preparation).
The cross correlation term will tend to reduce the dilution
effect of the dusty-SF EROs to the angular clustering of all EROs.  }
In any case, although for $z\sim1$ early-type galaxies the 
spatial clustering amplitude of $r_0=12\pm3$ \h1 Mpc
(derived in D01) is more secure, being based on a relatively large sample,
and consistent with the present analysis, it is likely that
such amplitude should be revised upward in light of the findings presented 
here.

\subsection{Analysis of systematic effects}

We tested the stability of these results 
with respect to the statistical uncertainty in the shapes of the selection
functions, which mostly influences the numbers of widely separated pairs.
A change in the pairwise peculiar velocity dispersion 
$\sigma_{12}$ by 20\%, would result
in a change of only about 10\% for the estimated $r_0$ values,
influencing of course the analysis of both ERO species in the same
direction and thus leaving the result unchanged.
Fig. \ref{fig:huub} shows that the two closest pairs
for the old EROs in our survey are expected at $\simlt5$ \h1 
Mpc separation at the $2\sigma$ level for $r_0\sim10$ \h1 Mpc, 
thus demonstrating that our result would hold correctly even if,  
because of redshift errors and roundings, the two closest pairs had been
found at $\Delta z=0.001$--0.002.
The effect of redshift errors is in fact negligible for
the dusty-SF ERO pairs, being all of them at $>20$ \h1 Mpc
separation.
Finally, we tested that the result is stable to variations of the 
color threshold at least up to $R-K>4.5$.

\section{Discussion}
\subsection{The clustering of dusty-SF EROs}

The clustering of 
dusty-SF EROs is small, and maybe consistent with the values of $1\simlt
r_0/($\h1 Mpc$) \simlt 2.5$ measured for star-forming galaxies at
$z\sim1$ (Le Fevre et al. 1996, Carlberg et al. 1997, Hogg et al.
2000). This would suggest the former to be a subclass of the latter,
but with stronger dust extinction. Locally, dusty galaxies detected by
IRAS are also known to have a relatively weak clustering 
(e.g. Saunders et al. 1992). 

The low level of clustering seems also to be
at odds with the idea that the dusty-SF EROs are
in a starburst phase following a major merger event, eventually expected to produce
an elliptical galaxy, as in this case one would expect to find a
correlation length somewhat lower than, but similar, to that of the
ellipticals at the same redshift.

SCUBA sub-mm selected sources are also thought to be dusty objects at 
high redshift
detected by virtue of the emission from dust warmed by star-formation
or AGN activity. This population is expected
(Magliocchetti et al. 2001) and tentatively observed (Scott et al. 2001)
to show strong angular clustering at the level of
$A(1^o)\sim0.01$ (see also Ivison et al. 2000 and Chapman et al. 2001).
Therefore our result suggest the dusty-SF
EROs are a different population with respect to SCUBA galaxies, 
with small overlap,
in agreement with the latter being typically fainter and more distant
($K>20$, median redshift $z\sim2.5$--3, {Smail et al. 2000}, 
; see also C02, Mohan et al. 2001, {and
Dannerbauer et al. 2002 for MAMBO sources}).


Finally, we note that the class of dusty-SF EROs could
be internally inhomogeneous: some of them maybe spirals with
moderate extinction (e.g. van Dokkum \& Stanford 2001) and there could
also be a mixture of dust-enshrouded AGNs and starburst galaxies (C02).

\subsection{Field $z\sim1$ early-type galaxies}

{Observations of samples of faint ERO galaxies have led to three key
conclusions regarding bright early-type galaxies with $L\simgt L_*$,
at $z\sim1$. Firstly, their space density is consistent with that of local
luminous early-type galaxies, when account is made of minimal pure
luminosity evolution (PLE) (C02). Secondly, spectroscopy implies age
$\simgt3$ Gyr for their stellar populations (assuming solar metalicity, C02); 
and thirdly, a comoving correlation length $r_0\simgt 12$ \h1
Mpc (this paper and D01) has been measured comparable with the
local value for luminous early-type galaxies.}

A large correlation length, $r_0 \simgt 10$ \h1 Mpc, is anticipated
theoretically for the hierarchical merging paradigm for which a
rapidly increasing bias is predicted for massive galaxies by $z\sim1$
(e.g. Mo \& White 1996, Moscardini et al. 1998).  {Such a large
correlation length is not expected for the PLE (galaxy conservation) 
scenario (D01).}
However, also current semi-analytical renditions of the 
hierarchical models seem to
be at odds with the observed results. For example, the Cole et
al. (2000) model predicts a comoving density (Fig. 1 of Benson et
al. 2001) of {\em all} the $z\sim1$ galaxies with $10^{11}M_\odot$
(consistent with our $K\leq19.2$ selection) which is a full order of
magnitude below the density of just the old EROs observed by C02.
Similarly, the Kauffmann et al. (1999) model
\footnote{http://www.mpa-garching.mpg.de/GIF/} predicts a comoving
density of $z\sim1$ EROs ($R-K\geq5$, $K\leq19.2$) that is 3(6) times
lower than observed by C02 for old(all) EROs.  In addition, in
these models $z\sim1$ galaxies qualified as field early-types appear
to have experienced recent star-formation, while the present sample of
old EROs is dominated by an old stellar population.  We conclude that
to our knowledge no semianalytical rendition of the hierarchical
merging models can yet account for all the 3 key observed properties
of $z\sim1$ field early type galaxies described above.

\begin{acknowledgements}
We thank the referee, Rob Ivison, for constructive remarks,
and Huub R\"ottgering and Sperello di Serego Alighieri 
for useful comments to the manuscript.
\end{acknowledgements}


\begin{thebibliography}{}

\bibitem[2001]{bem}{Benson A.J., Ellis R.S. \& Menanteau F., 2001, MNRAS, submitted (astro-ph/0110387)}
\bibitem[1997]{carl}{Carlberg R., Cowie L., Songaila A. \& Hu E., 1997, ApJ, 484, 538}
\bibitem[2001]{chap}Chapman S.C., Lewis G.F., Scott D., et al., 2001, ApJ 548, L17
\bibitem[1998]{cmn}{Cimatti A., Andreani P., R\"{o}ttgering H., Tilanus R., 1998, Nature 392, 895}
\bibitem[2001]{C02}{Cimatti A., Daddi E., Mignoli M., et al., 2002, A\&A,
381, L68 (C02)}
\bibitem[1999]{Co}{Cohen J.G., Hogg D.W., Pahre M.A. et al. 1999, ApJ, 120, 171}
\bibitem[2000]{Colee}{Cole S., Lacey C., Baugh C. \& Frenk C., 2000, MNRAS, 319, 168}
\bibitem[2000]{daddi2}{Daddi E., Cimatti A. \& Renzini A., 2000a, A\&A 362, L45}
\bibitem[2000]{daddi}{Daddi E., Cimatti A., Pozzetti L., et al., 2000b, A\&A 361, 535 (D00)}
\bibitem[2001]{d01}{Daddi E., Broadhurst T., Zamorani G., et al., 2001, A\&A, 376, 825 (D01)}
\bibitem[]{}{Dannerbauer H., Lehnert M., D. Lutz, et al., 2002, submitted to ApJ (astro-ph/0201104)}
\bibitem[2001]{X}{Firth A.E., Somerville R.S., McMahon R.G. et al. 2001, MNRAS, submitted (astro-ph/0108182)}
\bibitem[1990]{HSr}{Hartwick F.D. \& Schade D., 1990, ARA\&A, 28, 437}
\bibitem[2000]{hogg}{Hogg D., Cohen J. \& Blandford R., 2000, ApJ, 545, 32}
\bibitem[1999]{k99} {Kauffmann G., Colberg J.M., Diaferio A., White S.D.M., 1999, MNRAS 307, 529}
\bibitem[2000]{ivi}Ivison R.J., Dunlop J.S., Smail I., et al., 2000, ApJ, 542, 27
\bibitem[1998]{lsb}{Landy S., Szalay A. \& Broadhurst T., 1998, ApJ, 494, L133}
\bibitem[1996]{lef}{Le Fevre O., Hudon L., Lilly S., et al., 1996, ApJ, 461, 534}
\bibitem[2001]{maglio}{Magliocchetti M., Moscardini L., Panuzzo P., et al., 2001, MNRAS, 325, 1553}
\bibitem[2001]{} McCarthy P.~J., Carlberg R.~G., Chen H.-W., et al., 2001, ApJ,  560, L131
\bibitem[1996]{mw}{Mo H. \& White S.D.M., 1996, MNRAS, 282, 347}
\bibitem[2001]{mohan}{Mohan N., Cimatti A., R\"ottgering H., et al., 2002, A\&A in press (astro-ph/0112342)}
\bibitem[2000]{morio}{Moriondo G., Cimatti A. \& Daddi E., 2000, A\&A 364, 26}
\bibitem[1998]{mosca}{Moscardini L., Coles P., Lucchin F. \& Matarrese S., 1998, MNRAS 299, 95}
\bibitem[]{}Moustakas L.A. \& Somerville R.S., 2001, submitted to ApJ (astro-ph/0110584)
\bibitem[Peacock et al.~2001]{} Peacock J.~A., Cole S., Norberg P., et al., 2001, Nature,  410, 169
\bibitem[1992]{Sau}Saunders W., Rowan-Robinson M. \& Lawrence A., 1992, MNRAS, 258, 134
\bibitem[2001]{Scott}{Scott S., Fox M., Dunlop J., et al., 2001, submitted to MNRAS (astro-ph/0107446)}
\bibitem[1984]{shaver}{Shaver P., 1984, A\&A, 136, L9}
\bibitem[]{}Smail I., Ivison R.J., Kneib J.-P., et al., 1999, MNRAS, 308, 1061
\bibitem[]{} Smail I., Ivison R.~J., Owen F.~N., et al., 2000, ApJ,  528, 612
\bibitem[2001]{Smk}{Smith G.P., Smail I., Kneib J.-P., et al. 2001, MNRAS, in press (astro-ph/0109465)}
\bibitem{sp1}{Soneira S., Peebles P.J.E., 1977, ApJ 211, 1}
\bibitem{sp2}{Soneira S., Peebles P.J.E., 1978, AJ 83, 845}
\bibitem[1999]{spinrad}{Spinrad H., Dey A., Stern D., et al., 1997, ApJ 484, 581}
\bibitem[2000]{sttr}{Stiavelli M. \& Treu T., 2000, in
"Galaxy Disks and Disk Galaxies", ASP conference series
(astro-ph/0010100)}
\bibitem[2001]{2001JETPL..74....1T} Tinyakov P.~G., Tkachev I.~I., 2001, JETP Letters,  74, 1
\bibitem[2001]{vdk}{van Dokkum P. \& Stanford S., 2001, ApJ, 562, L35}

\end{thebibliography}
\end{document}